\documentclass[a4paper]{article}
\usepackage{ISCSLP2026}
\usepackage{ifthen}
\newboolean{blind}
\setboolean{blind}{false} 
\usepackage{listings}
\usepackage{xcolor}
\usepackage{tabularx}
\usepackage{enumitem}
\usepackage{xurl}
\usepackage[colorlinks=true, linkcolor=blue, citecolor=blue, urlcolor=blue]{hyperref}
\lstdefinestyle{jsonstyle}{
  basicstyle=\ttfamily\scriptsize,
  frame=single,
  breaklines=true,
  columns=fullflexible,
  showstringspaces=false,
  xleftmargin=1mm,
  xrightmargin=1mm
}
\title{ISCSLP 2026 CoT-TTS Challenge: Chain-of-Thought Reasoning for Context-Aware Text-to-Speech}

\name{
	\ifthenelse{\boolean{blind}}{Anonymous to ISCSLP}
	{Wei~Xue$^1$, Junlan~Feng$^2$, Shilei~Zhang$^3$, Yue~Wang$^4$, Ruosong~Yang$^4$, Bei~Liu$^1$, Liumeng~Xue$^5$, Sitong~Cheng$^1$, Jiahao~Pan$^1$, Weizhen~Bian$^1$, Boyi~Kang$^1$, Bin~Long$^6$}
}

\address{
	\ifthenelse{\boolean{blind}}{Anonymous to ISCSLP}
	{
		$^1$The Hong Kong University of Science and Technology\\
		$^2$China Mobile\\
		$^3$Jiutian Artificial Intelligence Technology (Beijing) Co., Ltd., China Mobile\\
		$^4$China Mobile (Hong Kong) Innovation Research Institute\\
		$^5$Nanjing University\\
		$^6$Hong Kong Generative AI Research \& Development Center
	}
}

\email{
	\ifthenelse{\boolean{blind}}{Anonymous to ISCSLP}
	{}
}

\begin{document}

\maketitle

\begin{abstract}
Recent advances in text-to-speech (TTS) have greatly improved speech naturalness, speaker similarity, and controllability. However, most existing controllable TTS systems still rely on explicit user-provided style prompts, making it difficult to automatically determine how a sentence should be spoken in long and complex conversational scenarios. This proposal introduces the ISCSLP 2026 CoT-TTS Challenge, which aims to evaluate whether a system can infer the intended speaking manner from contextual information and generate speech consistent with both the reasoning output and the surrounding scene. The challenge contains two tracks: text-context-aware CoT-TTS and audio-context-aware CoT-TTS. We construct a large-scale bilingual training set from speech-rich media and provide carefully filtered evaluation data for leaderboard comparison. Each system is required to output both a chain-of-thought reasoning analysis and the generated speech waveform. The official evaluation combines objective metrics, multimodal LLM-based evaluation, and human subjective assessment. To facilitate reproducibility, we provide inference code together with a fine-tuning recipe for a 0.6B Qwen3-based model trained via a three-stage strategy. This challenge is expected to support research on context understanding, chain-of-thought reasoning, and expressive speech generation for applications such as film dubbing, audiobook production, virtual characters, and spoken dialogue agents. Further information about the associated challenge is available at: \href{https://iscslp2026-cot-tts.github.io/challenge-website/}{\url{https://iscslp2026-cot-tts.github.io/challenge-website/}}
\end{abstract}
\noindent\textbf{Index Terms}: CoT-TTS, Chain-of-Thought Reasoning, Context-Aware Text-to-Speech, Expressive Speech Synthesis, Speech Foundation Models.

\section{Session/Challenge Organizers}

The proposed session organizers are listed below, together with their contact information and brief biographies where available for reference and further communication.\\

\noindent\textbf{Wei Xue}, The Hong Kong University of Science and Technology\\
\textit{Email:} weixue@ust.hk\\
\textit{Brief biography:} Wei Xue is an Assistant Professor at the Hong Kong University of Science and Technology. His research interests include audio processing, AI music, foundation models, generative AI, and multimodal audio understanding and generation.\\

\noindent\textbf{Junlan Feng}, China Mobile\\
\textit{Email:} fengjunlan@cmjt.chinamobile.com\\
\textit{Brief biography:} Junlan Feng is an IEEE Fellow and Chief Scientist at China Mobile. Her research interests include artificial intelligence, big data, and AI technologies for the communications industry, with a particular focus on the development of the Jiutian AI Platform.\\

\noindent\textbf{Shilei Zhang}, Jiutian Artificial Intelligence Technology (Beijing) Co., Ltd., China Mobile\\
\textit{Email:} zhangshilei@cmjt.chinamobile.com\\
\textit{Brief biography:} Shilei Zhang is the speech technology lead at China Mobile Jiutian Research. His research interests include speech recognition, voiceprint recognition, and speech large language models, with a particular focus on speech technology development for the Jiutian AI Platform.\\

\noindent\textbf{Yue Wang}, China Mobile (Hong Kong) Innovation Research Institute\\
\textit{Email:} yuewang@cmi.chinamobile.com \\
\textit{Brief biography:} Yue Wang is an AI Technical Research Senior Executive at China Mobile (Hong Kong) Innovation Research Institute. Her research interests include speech-to-speech translation and numerical optimization algorithms for large language models, with a particular focus on audio translation and efficient LLM optimization.\\

\noindent\textbf{Ruosong Yang}, China Mobile (Hong Kong) Innovation Research Institute\\
\textit{Email:} yangruosong@cmi.chinamobile.com\\
\textit{Brief biography:} Ruosong Yang is an AI Technical Research Senior Executive at China Mobile (Hong Kong) Innovation Research Institute. His research interests include large language models and multi-agent systems, with a particular focus on LLM-based multi-agent applications.\\

\noindent\textbf{Bei Liu}, The Hong Kong University of Science and Technology\\
\textit{Email:} beiliu@ust.hk\\
\textit{Brief biography:} Bei Liu is currently a Postdoctoral Fellow at The Hong Kong University of Science and Technology. His research interests include spoken dialogue systems, model compression, and inference optimization, with a particular focus on efficient speech and language modeling.\\

\noindent\textbf{Liumeng Xue}, Nanjing University\\
\textit{Email:} lmxue@nju.edu.cn\\
\textit{Brief biography:} Liumeng Xue is an Assistant Professor at Nanjing University. Her research interests include speech, music, and general audio understanding and generation, with a particular focus on controllable speech generation, emotional and expressive speech generation.\\

\noindent\textbf{Sitong Cheng}, The Hong Kong University of Science and Technology\\
\textit{Email:} schengaq@connect.ust.hk\\
\textit{Brief biography:} Sitong Cheng is a Ph.D. student at The Hong Kong University of Science and Technology. His research focuses on text-to-speech synthesis and speech-to-speech translation, with the goal of enabling more natural, seamless, and expressive voice communication.\\

\noindent\textbf{Jiahao Pan}, The Hong Kong University of Science and Technology\\
\textit{Email:} jpanbb@connect.ust.hk\\
\textit{Brief biography:} Jiahao Pan is a Ph.D. student at The Hong Kong University of Science and Technology. His research focuses on audio and music understanding and generation, covering music generation, speech enhancement, music separation, and multimodal audio-visual generation. 
\\

\noindent\textbf{Weizhen Bian}, The Hong Kong University of Science and Technology\\
\textit{Email:} wbian@connect.ust.hk\\
\textit{Brief biography:} Weizhen Bian is a Ph.D. student at The Hong Kong University of Science and Technology. His research focuses on text-to-speech synthesis, speech generation, and spoken language processing.\\

\noindent\textbf{Boyi Kang}, The Hong Kong University of Science and Technology\\
\textit{Email:} bkangaa@connect.ust.hk\\
\textit{Brief biography:} Boyi Kang is a Ph.D. student at Hong Kong University of Science and Technology. His research interests include speech, music, and general audio understanding and generation and multimodal intelligence.\\

\noindent\textbf{Bin Long}, Hong Kong Generative AI Research \& Development Center\\
\textit{Email:} lblongbin@163.com\\
\textit{Brief biography:} Bin Long is a backend engineer at Hong Kong Generative AI Research \& Development Center. His work focuses on scalable AI model service infrastructure, with a particular focus on unified access and deployment for LLM, TTS, ASR, and other speech and language model services.

\section{Background, Motivation, and Potential Impact}

Recent advances in large language models have expanded the capabilities of text-to-speech (TTS) systems~\cite{xie2025towards}. Modern TTS models can now support expressive style control, instruction-following generation, and long-form multi-speaker speech synthesis. Natural-language style prompts have been used for expressive TTS control~\cite{yang2024instructtts}, and recent systems further improve control over speaker timbre, emotion, duration, and speaking style~\cite{liao2026fish,zhou2025voxcpm,zhou2026indextts2}. Long-form speech generation systems also cover scenarios such as podcasts, multi-speaker conversations, and narrative audio generation~\cite{song2026borderless,xie2025soulx}. Despite these advances, most controllable TTS systems still rely on explicit user instructions or style prompts. In real-world use, such as conversations, film dubbing, and virtual characters, setting the speaking style for each sentence by hand is time-consuming and impractical. This motivates CoT-TTS, where the model uses the preceding context to reason about how the sentence should be spoken, produces an explicit chain-of-thought analysis, and generates speech that matches the scene.\\

\noindent A simple way to build CoT-TTS is to use a cascaded pipeline, such as ASR-LLM-TTS~\cite{yang2026duplexcascade} or AudioLLM-TTS~\cite{li2024style}. However, this design can introduce extra latency and information loss between modules. The final speech is also limited by the ability of the downstream TTS model. Recent studies have therefore tried to use historical dialogue context directly for conversational speech synthesis. They use text, audio, or multi-modal context to improve prosody, naturalness, and contextual appropriateness~\cite{xue2023m,seong2024h4c,wu2025diffcss,deng2023cmcu}. However, the link between context understanding and speech style control is usually implicit. As a result, it is hard to inspect why a certain speaking style is selected, or to evaluate whether the chain-of-thought reasoning is reasonable. ActorMind~\cite{chen2026actormind} introduces a multi-agent, CoT-style framework for speech role-playing. However, it mainly focuses on role-playing with predefined role profiles, which is different from synthesizing a user-specified target sentence with a given reference speaker timbre. Commercial speech LLMs, such as GPT-4o, have also demonstrated strong capabilities in context-aware spoken interaction~\cite{lin2025preliminary,hurst2024gpt}. However, these systems are typically proprietary and offer limited control over the target speaker timbre. CapTalk~\cite{su2026captalk} is the most related work to our setting. However, its control signal is mainly organized as caption-style or attribute-level descriptions. Its reasoning process is not explicitly evaluated as an output that explains why the target sentence should be spoken in a certain way. These limitations highlight the need for a benchmark that requires systems to both produce explicit chain-of-thought reasoning about the intended speaking manner and generate speech consistent with that reasoning.\\

\noindent Data availability is another key challenge for CoT-TTS. Many conversational TTS studies are built on DailyDialog~\cite{li2017dailydialog} or DailyTalk~\cite{lee2023dailytalk}.
DailyDialog is a multi-turn dialogue dataset, but it mainly contains daily-life conversations and does not provide speech recordings. DailyTalk records 2,541 dialogues sampled and modified from DailyDialog, making it a useful spoken dialogue dataset for conversational TTS. However, because it still inherits the daily-conversation nature of DailyDialog, its speaker diversity and scenario complexity remain limited. Later works further extend DailyTalk with emotional annotations or LLM-generated prompts, such as emotion categories, emotion intensity, and rationales~\cite{liu2024emotion,jeon2025prompt}. These annotations improve controllability, but the underlying dialogue scenarios are still relatively simple and the overall data scale remains limited. To obtain richer conversational scenes, some recent datasets are collected from films, TV shows, or situated dialogue settings. For example, NCSSD collects natural conversational speech and TV-show dialogues, with about 236 hours of Chinese and English data~\cite{liu2024generative}. EmoSpeech also constructs emotionally rich speech annotations from media speech segments, but its scale is still limited, with about 16 hours of audio~\cite{bian2024emospeech}. DNASpeech provides contextualized speech with dialogues, narratives, and actions, but its speech duration is only about 18 hours~\cite{cheng2025dnaspeech}. ActorMind introduces speech role-playing data with role and scene information, but the benchmark contains a limited number of roles and scenes~\cite{chen2026actormind}. This motivates the construction of a larger dataset with chain-of-thought reasoning annotations for CoT-TTS.\\

\noindent To address these challenges, we propose the ISCSLP 2026 CoT-TTS Challenge. This challenge evaluates whether a system can infer the intended speaking manner from textual or acoustic context, produce an explicit chain-of-thought reasoning process, and generate speech that matches both the reasoning output and the surrounding scene. We provide two tracks, a bilingual training set, a unified evaluation protocol, and a reproducible baseline to support comparison of different modeling strategies. The proposed challenge is expected to have impact on both research and applications. From a research perspective, it provides a benchmark for studying how context-aware TTS models understand dialogue history, infer speaker intention, track emotional change, and convert reasoning into acoustic realization. From an application perspective, CoT-TTS is related to film dubbing, virtual characters, and spoken dialogue systems, where the same sentence may need to be spoken in different ways depending on the scene. By making the reasoning process an explicit part of the task and evaluation, this challenge can help promote more controllable, explainable, and context-aware TTS systems.

\section{Tentative List of Authors / Teams Who Could Contribute Papers}

We expect this challenge to attract participating teams from both academia and industry. The topic is closely related to current research in speech synthesis, speech foundation models, spoken dialogue systems, and multimodal speech processing. The following list represents potential contributing communities and outreach targets, rather than confirmed submissions. Based on preliminary interest and the relevance of existing research directions, a tentative list of potential participating teams is shown in Table~\ref{tab:contributors}.

\begin{table}[h]
\caption{Tentative list of potential participating teams.}
\label{tab:contributors}
\centering
\small
\begin{tabular}{l}
\toprule
\textbf{Potential Participating Team} \\
\midrule
Tsinghua University \\
Tencent \\
Nanjing University \\
Huawei \\
The Chinese University of Hong Kong \\
Institute of Acoustics, Chinese Academy of Sciences \\
The Hong Kong University of Science and Technology \\
Institute of Artificial Intelligence (TeleAI) \\
Shanghai Jiao Tong University \\
National University of Singapore \\
JD.COM \\
University of Science and Technology of China \\
Nanyang Technological University \\
Beijing University of Posts and Telecommunications \\
Fudan University \\
\bottomrule
\end{tabular}
\end{table}

\section{Tentative Challenge Timeline}

The tentative timeline is designed to follow the official ISCSLP 2026 schedule. The challenge will be launched after proposal acceptance, and the model submission deadline will be aligned with the paper submission deadline. The evaluation stage will take place before the paper acceptance notification, so that the final rankings can be announced together with the accepted challenge papers. The specific schedule is shown in Table~\ref{tab:timeline}.\\

\begin{table}[h]
\caption{Tentative challenge timeline.}
\label{tab:timeline}
\centering
\begin{tabular}{ll}
\toprule
\textbf{Milestone} & \textbf{Date} \\
\midrule
Challenge Proposal Notification & June 15, 2026 \\
Challenge Website Launch & June 20, 2026 \\
Dataset and Baseline Release & June 20, 2026 \\
First Model Submission Round & July 10--18, 2026 \\
Final Model Submission Round & July 19--25, 2026 \\
Model Submission Deadline & July 25, 2026 \\
Paper Submission Deadline & August 3, 2026 \\
Objective Evaluation & July 26--August 8, 2026 \\
LLM-Based Evaluation & August 9--20, 2026 \\
Crowd-Sourced Human Evaluation & August 9--25, 2026 \\
Final Ranking and Winner Selection & August 26--30, 2026 \\
Paper Acceptance Notification & August 31, 2026 \\
Camera-Ready Deadline & September 21, 2026 \\
Challenge Session at ISCSLP 2026 & Conference Date \\
\bottomrule
\end{tabular}
\end{table}

\section{Planned Format of the Challenge Session}

The proposed challenge session will be organized as a two-hour event. The session will begin with an overview of the challenge, including the task setup, dataset construction, evaluation methodology, and overall results. Representatives from the winning teams in each track and model category will then be invited to present their approaches and share key findings. The session will conclude with a panel discussion on future directions of context-aware CoT-TTS and an award ceremony for the top-performing teams. A tentative schedule is shown in Table~\ref{tab:session_schedule}.\\

\begin{table}[h]
\caption{Tentative schedule of the challenge session.}
\label{tab:session_schedule}
\centering
\begin{tabular}{lr}
\toprule
\textbf{Activity} & \textbf{Duration} \\
\midrule
Opening Remarks and Challenge Overview & 10 min \\
Dataset and Baseline Introduction & 10 min \\
Evaluation Protocol and Results Summary & 10 min \\
Winner Presentation: Track 1, Unrestricted & 12 min \\
Winner Presentation: Track 1, Constrained & 12 min \\
Winner Presentation: Track 2, Unrestricted & 12 min \\
Winner Presentation: Track 2, Constrained & 12 min \\
Panel Discussion & 15 min \\
Audience Q\&A & 17 min \\
Award Ceremony and Closing Remarks & 10 min \\
\midrule
Total & 120 min \\
\bottomrule
\end{tabular}
\end{table}

\section{Recommended Expert Reviewers}

The following researchers are recommended as potential expert reviewers for papers submitted to this challenge session. They have relevant expertise in speech synthesis, speech generation, speech understanding, and spoken language processing.

\begin{table}[h]
\caption{Tentative list of recommended reviewers.}
\label{tab:reviewers}
\centering
\small
\begin{tabularx}{\linewidth}{@{}lX@{}}
\toprule
\textbf{Name} & \textbf{Affiliation} \\
\midrule
Xie Chen & Shanghai Jiao Tong University \\
Rui Liu & Inner Mongolia University \\
Berrak Sisman & Johns Hopkins University \\
Li Liu & The Hong Kong University of Science and Technology (Guangzhou) \\
Lei Xie & Northwestern Polytechnical University \\
Zhen-Hua Ling & University of Science and Technology of China \\
Xixin Wu & The Chinese University of Hong Kong \\
\bottomrule
\end{tabularx}
\end{table}

\section{Additional or Non-Standard Resources Needed}

\subsection{Training Dataset}

\begin{figure}[h]
  \centering
  \includegraphics[width=\linewidth]{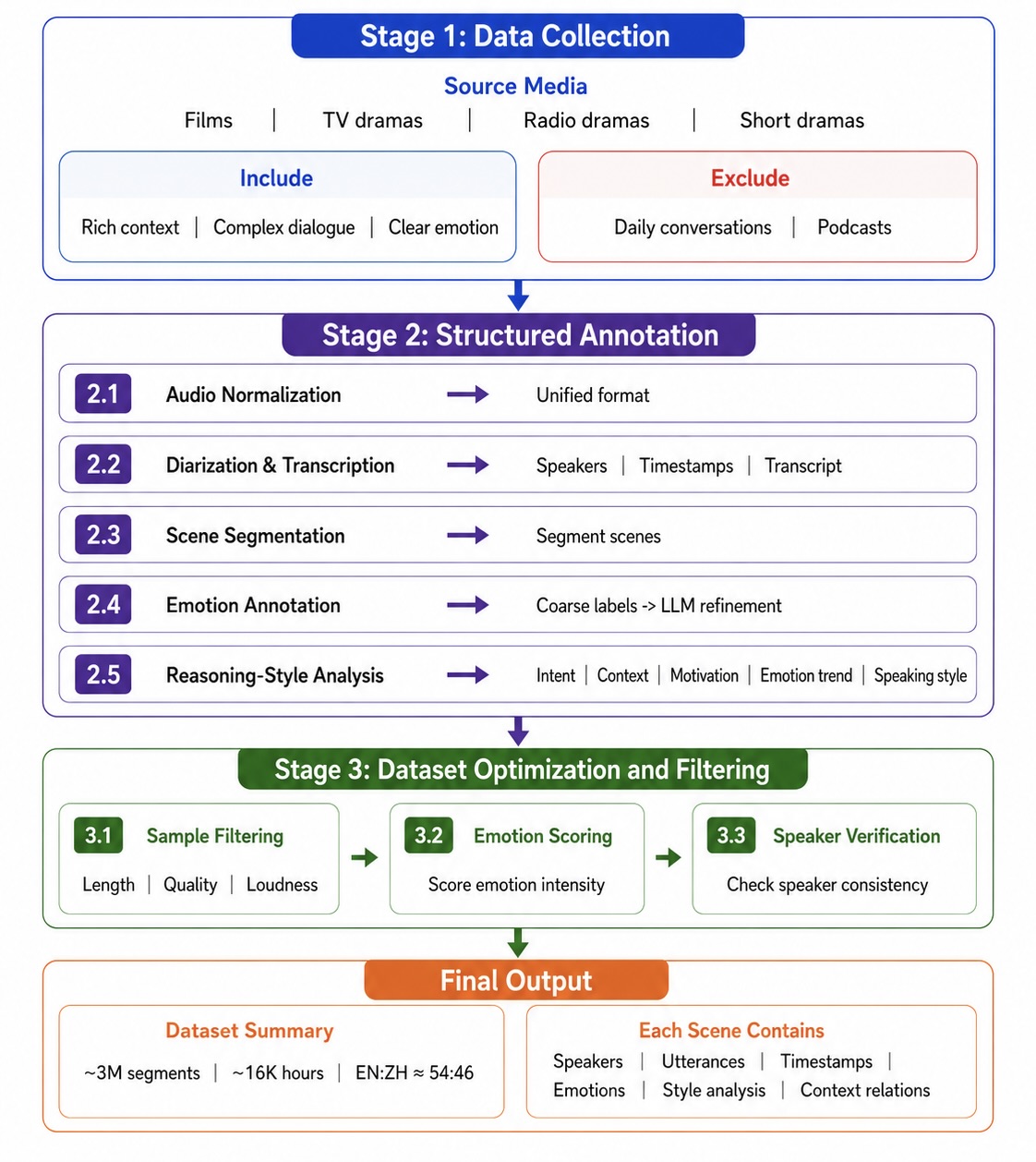}
  \caption{Overview of the initial training data construction pipeline. Raw media are converted and normalized, processed into utterance- and scene-level segments, annotated with emotion and reasoning information, and finally organized into CoT-TTS training samples.}
  \label{fig:data_pipeline}
\end{figure}

The training dataset is built from speech-rich media through a step-by-step processing pipeline. We first select suitable source materials and convert them into normalized audio. The audio is then processed to obtain utterance-level information such as speaker labels, timestamps, and spoken content. Based on these utterances, we further construct scene-level context and add emotion and chain-of-thought reasoning annotations. To ensure data quality, we filter the candidate samples according to contextual duration, effective speech duration, emotional intensity, loudness, and other audio-level features. We also use speaker-embedding similarity to reduce diarization-related errors~\cite{zeinali2019rvector,wang2023wespeaker}. Fig.~\ref{fig:data_pipeline} gives an overview of this initial construction process. \\

\noindent To build a dataset with rich conversational context, we collect data from films, TV dramas, radio dramas, and short dramas. These sources contain complex dialogue scenes and clear emotional changes, making them suitable for evaluating CoT-TTS. The copyright and data usage statement is provided in Appendix~\ref{app:copyright}. After collecting the source audio, we process it into structured utterance- and scene-level information. After audio conversion and normalization, we use the pyannote diarization model~\cite{Bredin23,Plaquet23} to obtain utterance-level information, including speaker labels, timestamps, and spoken content. Based on the timestamp distribution, each episode is roughly divided into candidate scenes. An LLM~\cite{deepseekai2025deepseekr1incentivizingreasoningcapability} then refines the scene boundaries and produces more coherent scene-level segments. For emotion annotation, we first obtain coarse emotion labels using FunASR~\cite{gao2023funasr}, which are then refined by the LLM according to the local scene context. In addition, each utterance is annotated by an LLM with a multi-dimensional reasoning analysis based on the preceding three to five utterances. The prompts used for LLM-based scene refinement and annotation are provided in Appendix~\ref{app:prompts}. As a result, each scene is represented as a structured segment containing speakers, utterances, timestamps, emotion annotations, reasoning analyses, and contextual relations among utterances.\\

\noindent After the initial construction, we refine the dataset through sample-level filtering. We extract audio-level features, including segment duration, effective speech duration, loudness, and emotion-related representations, to remove low-quality samples. For each audio segment, we obtain an emotion vector with arousal, dominance, and valence~\cite{wagner2023dawn}, and define the emotional expression intensity as
\begin{equation}
I_{\mathrm{emo}} = 0.7 \cdot A + 0.3 \cdot D,
\end{equation}
where $A$ and $D$ denote arousal and dominance, respectively. Segments with low emotional expression intensity are removed. Since pyannote-based diarization~\cite{Bredin23,Plaquet23} may be inaccurate in complex conversations, we further re-transcribe each segment using Qwen3-ASR~\cite{Qwen3-ASR} and perform speaker verification with speaker embeddings~\cite{zeinali2019rvector,wang2023wespeaker}. For each segment, we compare its embedding with five reference segments from the annotated speaker and keep the best match only when the cosine similarity is above 0.45. After filtering, the dataset contains around 3 million segments, corresponding to roughly 16K hours of speech, with an English-to-Chinese ratio of around 54:46. A summary of the released training data is shown in Table~\ref{tab:training_data_summary}.\\

\begin{table}[h]
\caption{Summary of the released training data.}
\label{tab:training_data_summary}
\centering
\small
\begin{tabular}{lccc}
\toprule
\textbf{Language} & \textbf{Duration} & \textbf{Ratio} & \textbf{Segments} \\
\midrule
English & $\sim$8.6K h & 54\% & $\sim$1.62M \\
Chinese & $\sim$7.4K h & 46\% & $\sim$1.38M \\
\midrule
Total & $\sim$16K h & 100\% & $\sim$3.0M \\
\bottomrule
\end{tabular}
\end{table}

\noindent The released audio files are extracted segments from the original source recordings. To reduce storage usage, the audio format is standardized to FLAC, while other acoustic properties are preserved as much as possible. The audio is not aggressively standardized or denoised; therefore, files may differ in sampling rate, channel configuration, loudness, background sounds, and environmental noise. This design preserves realistic acoustic conditions and allows participants to apply their own preprocessing, normalization, enhancement, or filtering methods. The metadata and annotations are generated from normalized and denoised versions of the corresponding segments to improve annotation reliability. The released training dataset is available at: \url{https://huggingface.co/datasets/HKUSTAudio/ISCSLP2026-CoT-TTS}\\

\noindent The released dataset is organized into language- and partition-specific folders. Each folder contains one metadata JSON file and two audio directories: \texttt{dialogue\_segments/} and \texttt{continuous\_segments/}. The \texttt{dialogue\_segments/} directory contains utterance-level clips extracted according to diarization timestamps, including target speech, reference speech, and dialogue-context speech. In contrast, \texttt{continuous\_segments/} contains the corresponding clips with extended boundaries, so that the original acoustic continuity of the dialogue context is better preserved. This design allows users to choose between cleaner utterance-level segments and more context-preserving continuous audio segments. The detailed data structure is provided in Appendix~\ref{app:initial_data_format}.

\subsection{Baseline}

To demonstrate the feasibility of the proposed task and its reproducibility on commonly available hardware, we fine-tune a 0.6B Qwen3-based baseline model on the released training data using a parameter-efficient adaptation strategy with RTX 4090 GPUs. The training strategy follows the three-stage training paradigm of UniSS~\cite{cheng2025uniss}. The results show that even under this lightweight setting, the baseline can produce meaningful reasoning analyses and demonstrates a certain degree of controllability in context-aware CoT-TTS. The baseline implementation is available on \url{https://github.com/iscslp2026-cot-tts/baseline}.\\ 

\noindent In the first stage, we perform modality alignment through ASR and TTS tasks. Text is tokenized into text tokens using the Qwen3 tokenizer~\cite{qwen3technicalreport}, while speech is encoded into audio tokens using BiCodec~\cite{wang2025spark}. For the ASR task, the model is trained to predict text tokens from input audio tokens. For the TTS task, the model takes text tokens and the global tokens of a reference speech sample as input and predicts both global and semantic tokens. In the second stage, we train the model on the main context-aware CoT-TTS task, where the input consists of either historical dialogue audio or historical dialogue text, the global tokens of the reference speech, and the target text. For the audio-context setting, the model first generates the transcription of the historical audio with emotion tags, then produces the reasoning analysis, and finally predicts the target audio tokens. For the text-context setting, the model directly generates the reasoning analysis and the target audio tokens. We also include auxiliary tasks to strengthen historical-audio understanding, speaker-aware transcription, audio feature analysis, and instruction-based TTS generation. We further mix the ASR and TTS tasks from the first stage to preserve the basic cross-modal alignment ability. In the third stage, we further fine-tune the model on a higher-quality subset of the training data. This final fine-tuning stage further enhances the stability of reasoning generation and style control in the generated speech.

\subsection{Challenge Website}

\noindent The website provides challenge descriptions, track definitions, dataset access, baseline code, submission instructions, important announcements, and evaluation details. The submission portal and public leaderboard will be added before the official submission period. The official challenge website is publicly available at:
\url{https://iscslp2026-cot-tts.github.io/challenge-website/}.

\section{Challenge Description}

This challenge focuses on context-aware CoT-TTS. Given contextual information, a reference speech sample, and a target text, a system is expected to infer the appropriate speaking style of the target sentence, produce a chain-of-thought reasoning analysis, and generate speech with the speaker timbre specified by the reference speech. Participants are required to build end-to-end systems, and cascaded systems such as ASR-LLM-TTS pipelines are not allowed. The challenge contains two tracks according to the type of context provided to the system: Track 1 uses textual context, while Track 2 uses acoustic context. For each track, we maintain two leaderboards: an unrestricted category and a parameter-constrained category. The parameter-constrained category is designed for systems with fewer than 1B parameters, making the challenge more accessible to teams with limited computational resources. In contrast, the unrestricted category allows larger models and supports comparison with large-scale systems.

\subsection{Track 1: Text-Context-Aware CoT-TTS}

Track 1 requires the system to generate contextually appropriate speech based on textual dialogue context. The input context consists of previous dialogue turns in the form of speaker-labeled transcripts, such as ``speaker-id: spoken text''. In addition to the dialogue history, the system is provided with the target text and a reference speech sample that specifies the target speaker timbre. Based on these inputs, the model should infer the appropriate speaking style, emotion, tone, and rhythm of the target utterance, and then generate speech that is both speaker-consistent and contextually coherent.\\

\noindent The textual contexts are collected from scenarios such as films, TV dramas, short dramas and radio dramas. These sources often contain conflict, tension, emotional changes, and narrative progression. Therefore, the model cannot rely only on the target sentence itself. It needs to use the preceding dialogue to infer the speaker's emotional state, communicative intention, and speaking style, and then produce speech that is coherent with the ongoing conversation.

\subsection{Track 2: Audio-Context-Aware CoT-TTS}

Track 2 uses the same general types of scenarios as Track 1, but the context provided to the system is changed from text to audio. Instead of reading previous dialogue in textual form, the model receives the preceding multi-speaker dialogue as a continuous audio segment, rather than as speaker-separated clips. The system is expected to infer the appropriate speaking style from the audio context and generate speech for the target text in a consistent and natural style.\\

\noindent This track is closer to real-world applications where previous speech is available but reliable transcripts or structured scene descriptions may not be provided. Given the continuous audio context, target text, and reference speech, the model should reason about how the target sentence should be spoken and generate speech that continues the previous context naturally.

\subsection{Input and Output Specifications}

Each test sample provides the contextual information, a target text, and a reference speech sample. The context is represented according to the corresponding track, while the reference speech is used only to specify the target speaker timbre. The system is required to output both a reasoning analysis describing the intended speaking manner of the target text and the generated speech waveform. The generated speech should preserve the reference speaker timbre and be consistent with the inferred speaking style. The input and output formats can be summarized as follows:\\

\noindent For Track 1, an example input can be represented as follows:
\begin{lstlisting}[style=jsonstyle]
  "context": ["speaker-0: I cannot believe this happened."...],
  "target_text": "Then we must act now.",
  "reference_audio": "ref_speaker.wav"
\end{lstlisting}

\noindent For Track 2, the textual context is replaced by a continuous audio context:
\begin{lstlisting}[style=jsonstyle]
  "context_audio": "history_dialogue.wav",
  "target_text": "Then we must act now.",
  "reference_audio": "ref_speaker.wav"
\end{lstlisting}

\noindent For both tracks, the required output includes a reasoning analysis and the generated speech waveform:
\begin{lstlisting}[style=jsonstyle]
  "reasoning": "The previous context indicates tension and urgency, so the target sentence should be spoken with a tense and decisive tone.",
  "output_audio": "generated.wav"
\end{lstlisting}

\subsection{Evaluation Data}

\begin{figure}[h]
  \centering
  \includegraphics[width=0.95\linewidth]{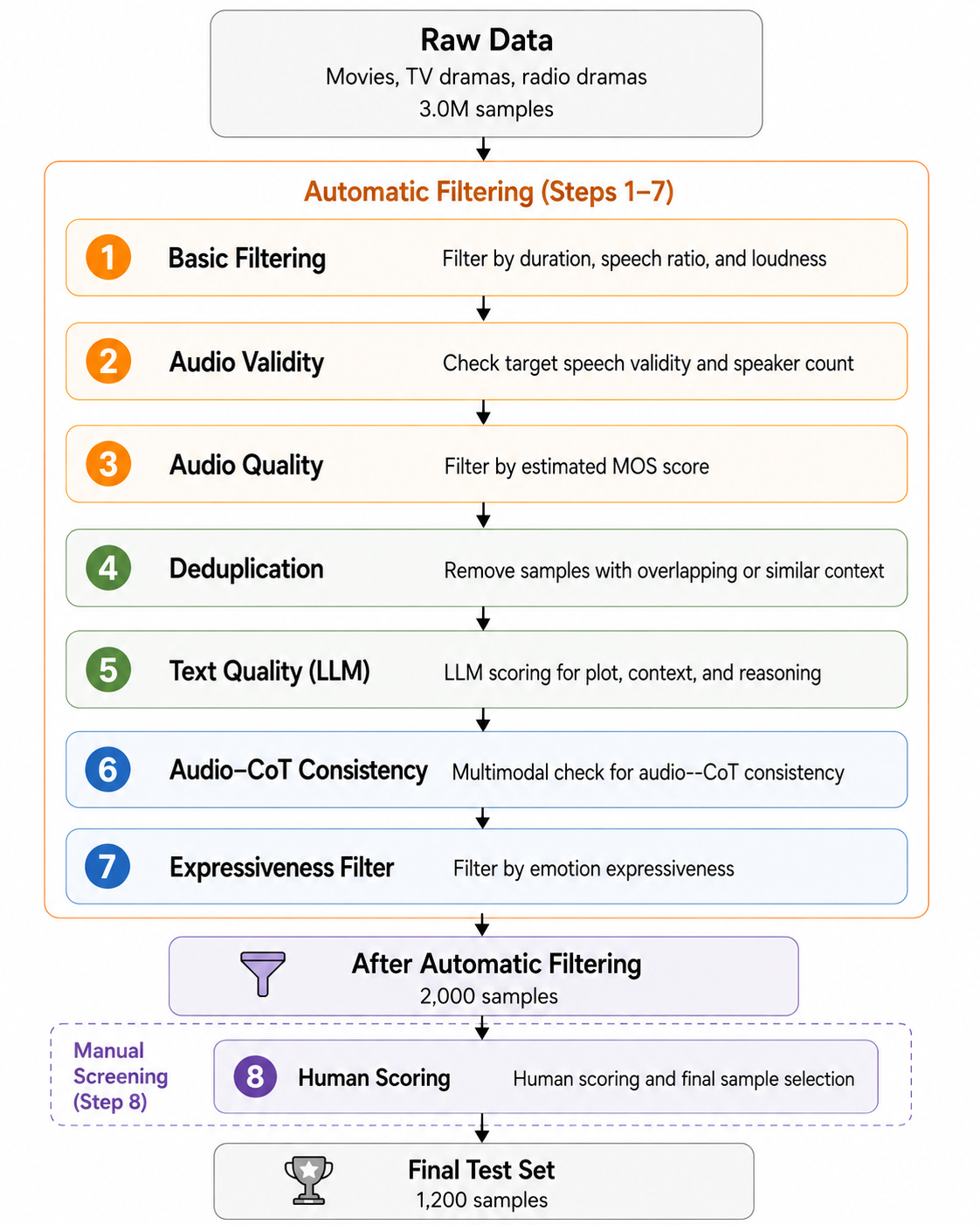}
  \caption{Overview of the evaluation data filtering pipeline. The initial candidate pool is progressively filtered through basic audio and context checks, target-audio quality assessment, similarity-based deduplication, LLM-based text quality scoring, audio--CoT consistency checking, and final human assessment.}
  \label{fig:eval_pipeline}
\end{figure}

To construct the evaluation set, we collect another pool of about 3 million samples from speech-rich media, with no overlap with the training data. The pool contains both Chinese and English samples and follows the same basic structure as the training set. Since the evaluation set is used for leaderboard comparison, we apply stricter filtering to ensure sufficient context, reliable annotations, and high-quality target speech. The detailed process is shown in Fig.~\ref{fig:eval_pipeline}.

\noindent The filtering process is conducted in multiple stages. We first remove samples with insufficient dialogue context, inappropriate audio duration, low effective speech ratio, weak emotional expressiveness, or unstable loudness. We then apply target-audio quality filtering based on estimated MOS scores~\cite{stahl2025distillation} and remove highly similar samples constructed from overlapping dialogue windows. To further improve the reliability of the evaluation set, we use a text-based LLM~\cite{deepseekai2025deepseekr1incentivizingreasoningcapability} to score the plot richness, context sufficiency, and correctness of the CoT reasoning analysis. In addition, an audio-capable multimodal model~\cite{qwen3technicalreport} is used to check whether the target speech is consistent with the reasoning summary, and samples with multiple speakers in the target audio are removed.\\

\noindent After automatic filtering, we further conduct human quality assessment through a web-based annotation interface. Annotators evaluate each sample from four aspects: transcription accuracy, scenario richness, CoT reasoning accuracy, and the consistency between the target audio and the reasoning analysis.  Samples with special cases, such as audiobook narration or multiple speakers in the target audio, are also excluded. The detailed web interface and scoring criteria are provided in Appendix~\ref{app:web_interface}. After this final manual filtering stage, the evaluation set contains 600 Chinese samples and 600 English samples. The selected samples cover diverse conversational scenarios with clear contextual dependencies, expressive speaking styles, and reliable target audio.

\subsection{Evaluation Method}

The official evaluation will be conducted by the organizers using a unified evaluation pipeline. The final score will consist of three parts: objective evaluation, LLM-based evaluation, and human subjective evaluation. The weights of these three parts will be 30\%, 20\%, and 50\%, respectively:

\begin{equation}
S = 0.3 S_{\mathrm{obj}} + 0.2 S_{\mathrm{LLM}} + 0.5 S_{\mathrm{human}},
\end{equation}

\noindent where $S_{\mathrm{obj}}$, $S_{\mathrm{LLM}}$, and $S_{\mathrm{human}}$ denote the normalized objective score, LLM-based score, and human subjective score.\\

\noindent Before detailed scoring, each submission will go through a basic validity check. Systems with invalid outputs, severe content mismatch, extremely low speech quality, or excessively high RTF may be excluded from later evaluation stages. This step is intended to ensure that subjective and LLM-based evaluation resources are used only on valid and comparable systems.\\

\noindent Objective evaluation will use TTS-related metrics to measure generated speech. Naturalness will be evaluated using UTMOSv2~\cite{baba2024utmosv2}, while speech quality will be measured using DNSMOS P.835~\cite{reddy2022dnsmos}. Intelligibility will be evaluated using CER or WER from automatic transcriptions~\cite{morris2004and}. Speaker similarity will be measured by extracting speaker embeddings from generated and reference speech, followed by cosine similarity~\cite{wang2023wespeaker}. Since each test sample is paired with ground-truth target speech, we will also consider prosody- and expression-related metrics, including F0 correlation, emotion expressiveness, and duration error. We will report real-time factor (RTF) to reflect inference efficiency. All objective metrics will be normalized and combined to obtain the final objective score $S_{\mathrm{obj}}$, which provides an estimate of system performance before human evaluation.\\

\noindent The LLM-based evaluation will focus on the reasoning output and its consistency with the generated speech. A fixed audio-capable multimodal LLM evaluator will be used to assess context understanding, internal reasoning coherence, and speech--reasoning consistency. In addition, we introduce a reasoning informativeness score to measure whether the reasoning output contains sufficiently specific and context-dependent analysis. This score is used as a modulation factor for the other LLM-based scores, so that overly generic reasoning with low information content cannot receive a high final LLM score even if it is superficially accurate. This design discourages trivial explanations, such as simply stating that a response is needed because someone has spoken, and encourages models to infer concrete communicative intent, emotional state, scene dynamics, and expected speaking style.\\

\noindent The human subjective evaluation will be conducted through a crowd-sourcing platform. To improve efficiency, each submitted model will be evaluated on a selected subset of the hidden test set, and each sample will be rated by at least five independent listeners. The final human score will be computed from valid ratings after quality control. Specifically, listeners will rate the generated results along four dimensions: coherence between the generated speech and the historical context, accuracy of the reasoning analysis, informativeness of the reasoning output, and consistency between the generated speech and the reasoning analysis. Scores from these dimensions will be aggregated to obtain the final human subjective score $S_{\mathrm{human}}$.\\

\subsection{Submission Requirements}

Participants are required to submit their systems through the official challenge website. Each submission should include the inference code, the runtime environment, the trained model or model checkpoints, and a short system description paper. The inference code should be directly executable on the evaluation set and should produce both the reasoning analysis and the generated speech waveform for each test sample. The runtime environment should include the necessary dependency information, such as a Docker file, environment file, or installation script, to ensure reproducibility. The submitted model files should contain all parameters required for inference, and the system description should briefly introduce the model architecture, training data, inference procedure, parameter scale, and any external resources used. These materials will be used to verify the validity of the submission and to support reproducible evaluation.

\subsection{Other Rules}

\begin{itemize}[leftmargin=*, itemsep=1pt, topsep=2pt, parsep=0pt]

    \item Participants may use the official training and development data released by the organizers.

    \item External datasets and publicly available pretrained models are allowed, provided that the datasets are publicly available and permitted for academic research. External data must not contain, reconstruct, or be derived from the hidden evaluation sources.

    \item Any external data, pretrained models, synthetic data, data augmentation strategies, or post-processing methods must be clearly described in the system description paper, which is required for final ranking and presentation.

    \item Cascaded systems, such as ASR--LLM--TTS pipelines, are not allowed.

    \item For the parameter-constrained category, the total number of parameters used during inference must be fewer than 1B. All models loaded or invoked during inference, including frozen modules, vocoders, speech tokenizers, speaker encoders, auxiliary models, and post-processing models, are counted toward the parameter budget.

    \item Manual inspection, use, or reverse engineering of the hidden test labels is not allowed. Participants must not attempt to identify, retrieve, reconstruct, or use the original source media corresponding to hidden evaluation samples.

    \item Reasoning outputs must follow the required format constraints. In particular, reasoning with low information content may receive a low reasoning-informativeness score even if it is superficially correct.

    \item The submitted inference code must run without internet access during official evaluation. All required model files, tokenizers, vocoders, and auxiliary resources must be included in the submission.

    \item Submissions whose real-time factor (RTF) exceeds 3.0, as measured by the organizers in the official evaluation environment, will not proceed to the official evaluation or final ranking.

\end{itemize}

\bibliographystyle{IEEEtran}

\bibliography{mybib}

\clearpage

\appendix

\appendix

\clearpage
\onecolumn
\section{Web Interface and Scoring Criteria for Human Quality Assessment}
\label{app:web_interface}

\begin{center}
\vspace{6.0em}

\includegraphics[width=0.95\textwidth]{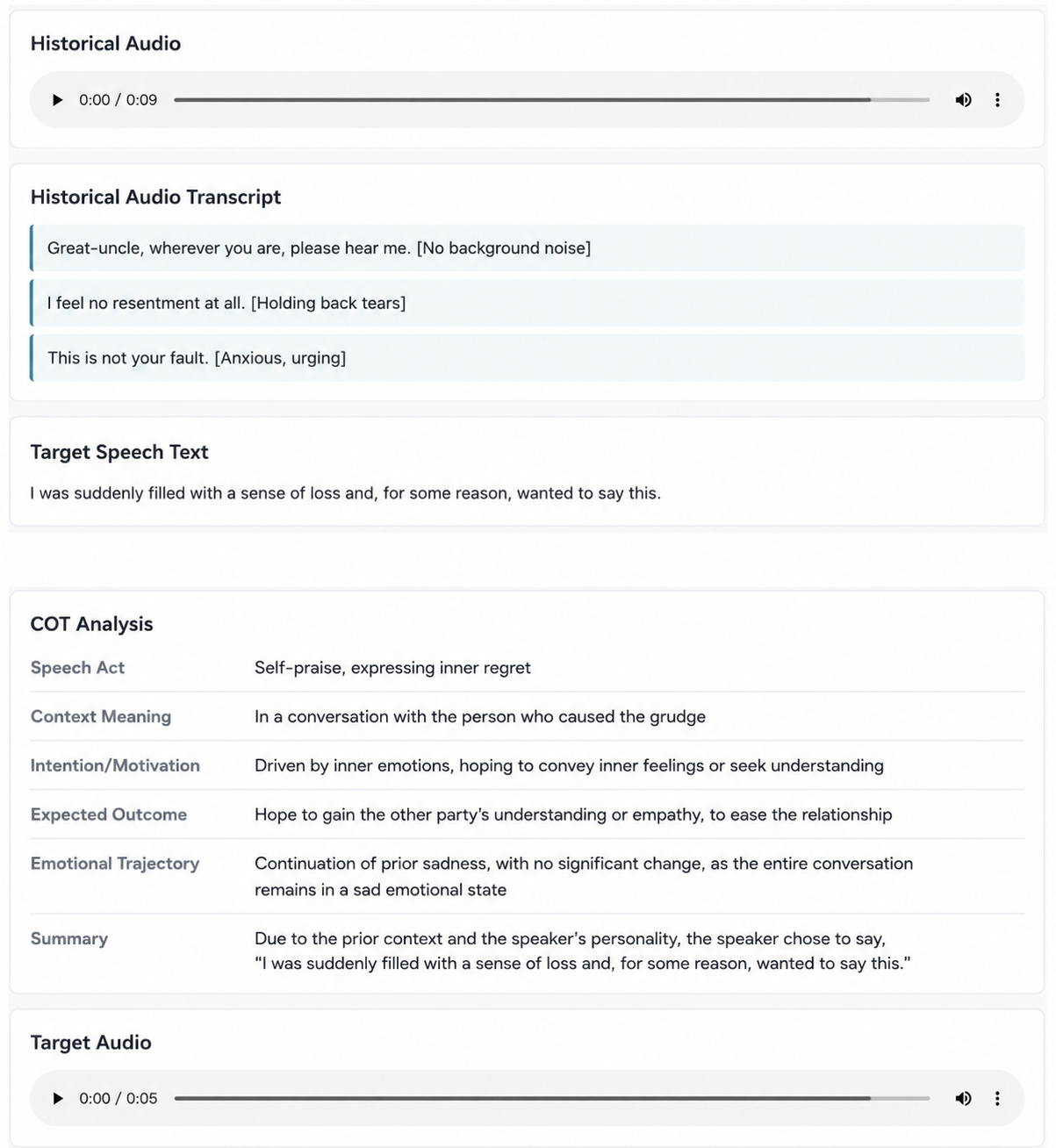}
\end{center}

\clearpage

\begin{center}
\vspace{7.0em}
\includegraphics[width=0.95\textwidth]{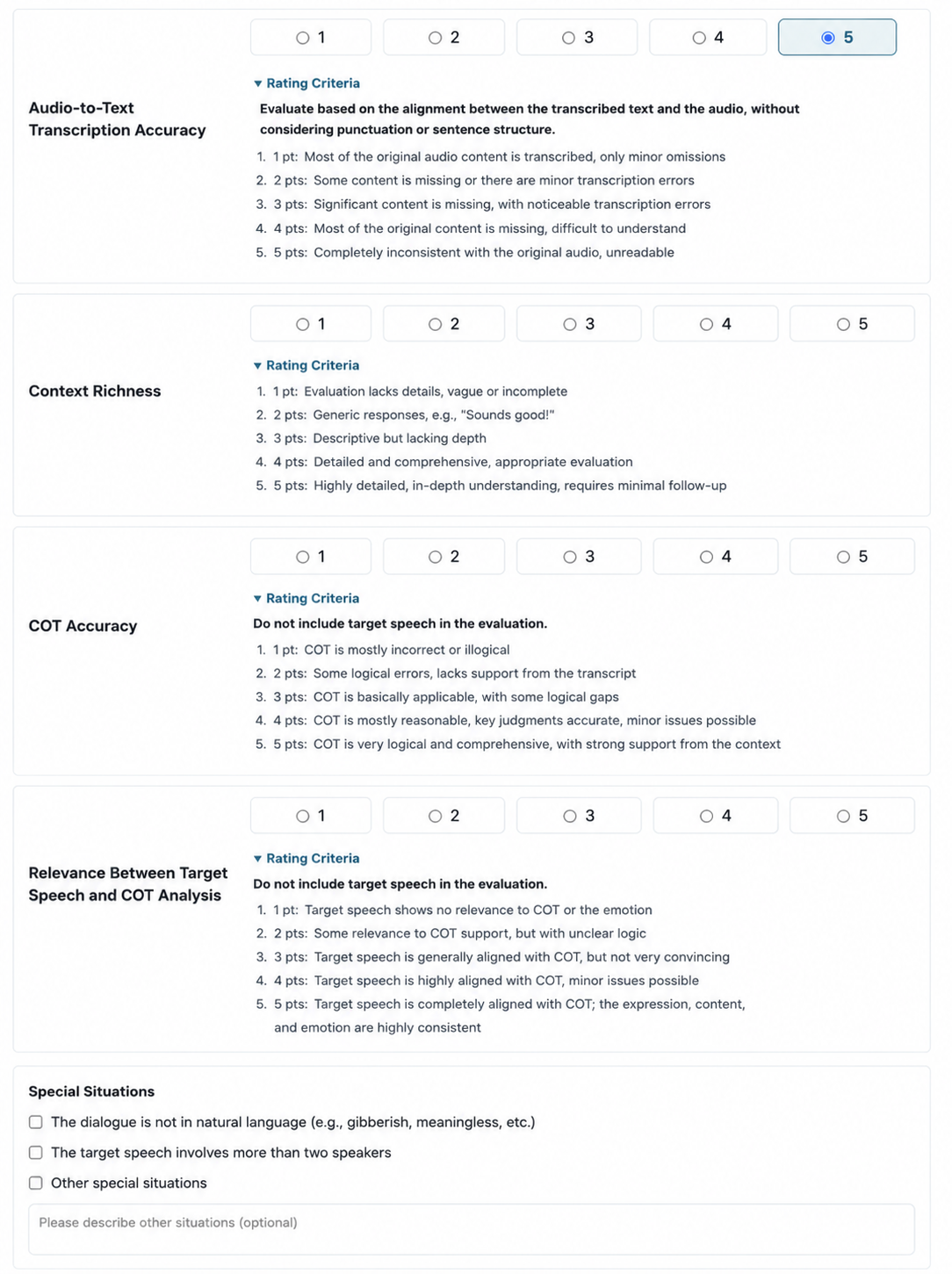}

\end{center}

\clearpage
\twocolumn

\section{Data Copyright Statement}
\label{app:copyright}

The challenge dataset is prepared for academic research and evaluation purposes only. The source materials are collected from publicly accessible media sources. The organizers do not claim ownership of the original media content, and the copyright of the original videos, audio programs, podcasts, films, dramas, or other source materials remains with their respective owners.

The released dataset will not include the original videos, complete audio programs, or full-length source materials. Instead, the source recordings will be processed through a data preparation pipeline, including speech segmentation, clipping, normalization, resampling, quality filtering, transcription, and annotation. Only the processed speech segments and the corresponding metadata or annotations will be released for the challenge.

The dataset is intended to support research on context-aware CoT-TTS. Participants may use the released data only for non-commercial research, development, and evaluation related to the challenge. Commercial use, redistribution of the dataset, and attempts to reconstruct, identify, or redistribute the original source materials are not permitted.

The dataset will be released under a non-commercial research-use license, following the spirit of the Creative Commons Attribution-NonCommercial 4.0 license. Participants are required to follow the data usage terms provided by the organizers. The organizers reserve the right to remove or modify any data item if copyright, privacy, or licensing concerns are raised.

\section{LLM Prompts for Data Construction}
\label{app:prompts}

\subsection{Scene Boundary Refinement}

\begin{lstlisting}[style=jsonstyle]
You are a scene editor for film and television dialogue. Given utterances with relative timestamps, identify scene split points based only on temporal continuity and topic consistency. A split index means that the utterance at this index starts a new scene. Do not include 0 or the total number of utterances. Return only a JSON array of split indices in ascending order.

Input:
- time gap threshold: {t_gap}
- block start time: {block_start}
- utterance list: {utterance_list}

Output example: [12, 37]
\end{lstlisting}

\subsection{Emotion Annotation}

\begin{lstlisting}[style=jsonstyle]
Given the dialogue window, target utterance, original emotion tag, and audio event information, generate a context-aware Chinese emotion tag.

The tag should be a short Chinese phrase describing emotional intensity, attitude, or speaking manner, rather than a bare emotion word. Use only the current dialogue window and make conservative judgments when the evidence is insufficient.

Input:
- target speaker: {speaker}
- target utterance: {text}
- original emotion tag: {raw_emo_tag}
- history note: {history_note}
- dialogue window: {window_lines}

Output:"emo_tag": "Chinese short phrase"
\end{lstlisting}

\subsection{Reasoning Analysis}

\begin{lstlisting}[style=jsonstyle]
Given the dialogue window, target utterance, refined emotion tag,
and audio event information, analyze the target utterance.
Use only the current dialogue window and output only a JSON object.

Input:
- target speaker: {speaker}
- target utterance: {text}
- refined emotion tag: {emo_tag}
- history note: {history_note}
- dialogue window: {window_lines}

Output:
{
  "dim1_language_act": {"need_cot": true, "text": "..."},
  "dim2_scene_semantics": {"need_cot": true, "text": "..."},
  "dim3_cognition_persona_motivation": {"need_cot": true, "text": "..."},
  "dim4_expected_outcome": {"need_cot": true, "text": "..."},
  "dim5_emotion_trajectory": {"need_cot": true, "text": "..."},
  "summ": "..."
}
\end{lstlisting}

\section{Processed Data Format}
\label{app:initial_data_format}

\begin{lstlisting}[style=jsonstyle]
    {
  "target_segment": {
    "segment_id": "Unique ID",
    "ref_segment_id": "Segment ID",
    "text": "Text content",
    "speaker": "Original speaker label",
    "normalized_speaker": "Normalized speaker label",
    "emotion_tag": "Emotional description",

    "features": {
      "duration": "Total duration",
      "active_duration": "Valid duration",
      "loudness": "Loudness of the audio",
      "expressive_intensity": "Expressive intensity"
    },
    "cot_text": "Chain-of-thought style analysis",
    "start": "Start time",
    "end": "End time",
    "best_ref_similarity": "Similarity"
  },
  "dialog_segments": [
    {
      "segment_id": "Unique ID",
      "text": "Text content",
      "speaker": "Original speaker label",
      "normalized_speaker": "Normalized speaker label",
      "emotion_tag": "Emotional description",
      "features": {
      "duration": "Total duration",
      "active_duration": "Valid duration",
      "loudness": "Loudness of the audio",
      "expressive_intensity": "Expressive intensity"
     },

      "start": "Start time",
      "end": "End time"
    },
      ...
  ],
}

\end{lstlisting}

\end{document}